\documentclass[12pt,preprint]{aastex}
\usepackage[dvips]{color}

\def\ltsima{$\;\buildrel < \over \sim \;$}
\def\simlt{\lower.5ex \hbox{\ltsima}}
\def\gtsima{$\;\buildrel > \over \sim \;$}
\def\simgt{\lower.5ex \hbox{\gtsima}}
\newcommand{\asec}{$^{\prime\prime}$}
\newcommand{\amin}{$^{\prime}$}

\newcommand{\re}{\textcolor[rgb]{0,0,0}}
\newcommand{\ree}{\textcolor[rgb]{0,0,0}}

\vfill
\eject\shorttitle{Warm H$_2$ with IRAC}
\shortauthors{Neufeld and Yuan}

\begin{document}

\title{Mapping warm molecular hydrogen with \\
{\it Spitzer's} Infrared Array Camera (IRAC)}
\author{David A. Neufeld and Yuan Yuan}
\affil{Department of Physics and Astronomy, Johns Hopkins University, 3400 North Charles Street, Baltimore, Maryland}

\begin{abstract}

\re{Photometric maps, obtained with {\it Spitzer}'s Infrared Array Camera (IRAC), can provide a valuable probe of warm molecular hydrogen within the interstellar medium.}  IRAC maps of the supernova remnant IC443, extracted from the {\it Spitzer} archive, are strikingly similar to spectral line maps of the H$_2$ pure rotational transitions that we obtained with the Infrared Spectrograph (IRS) instrument on {\it Spitzer}.   IRS spectroscopy indicates that IRAC Bands 3 and 4 are indeed dominated by the H$_2$ $v=0$--0 S(5) and S(7) transitions, respectively.  Modeling of the H$_2$ excitation suggests that Bands 1 and 2 are dominated by H$_2$ $v = 1-0$ O(5) and $v=0$--0 S(9).   Large maps of the H$_2$ emission in IC433, obtained with IRAC, show band ratios that are inconsistent with \re{the presence of gas at a single temperature.}  The relative strengths of IRAC Bands 2, 3, and 4 are consistent with pure H$_2$ emission from shocked material with a power-law distribution of gas temperatures.  CO vibrational emissions \ree{do not} contribute significantly to the observed Band 2 intensity.
Assuming that the column density of H$_2$ at temperatures $T$ to $T+dT$ is proportional to $T^{-b}$ for temperatures up to 4000~K, we obtained a typical estimate of 4.5 for $b$.  The power-law index, $b$, shows variations over the range $\sim 3 - 6$ within the set of different sight-lines probed by the maps, with the majority of sight-lines showing $b$ in the range $4-5$.  The observed power-law index is consistent with the predictions of simple models for paraboloidal bow shocks.

\end{abstract}


\section{Introduction}

Warm molecular hydrogen, at temperatures of several hundred to several thousand K, has been observed widely in molecular clouds that have been heated by interstellar shock waves (e.g. Gautier et al.\ 1976, Treffers et al.\ 1976, Combes \& Pineau des Forets 2000.)  Such clouds are frequently associated with protostellar outflows and supernova remnants (SNR), and emit a rich spectrum of rovibrational and pure rotational H$_2$ emissions.  While the near-infrared rovibrational transitions have been observed extensively with ground-based observatories, observations of the mid-infrared pure rotational transitions are significantly hampered by atmospheric absorption; thus ground-based observations can provide only a fragmentary picture of the pure rotational spectrum of warm H$_2$, and space-based observations from the {\it Infrared Space Observatory (ISO)} or the {\it Spitzer Space Telescope} have been needed to provide a complete spectrum.  In particular, recent observations with the Infrared Spectrograph (IRS) on {\it Spitzer} have provided sensitive spectral line maps for the S(0) through S(7) pure rotational transitions of H$_2$, allowing the gas temperature and ortho-to-para ratio of the emitting H$_2$ to be determined.  Such observations have revealed the presence of an admixture of multiple gas temperatures along every sight-line that has been observed, with non-equilibrium ortho-to-para ratios readily apparent in the cooler gas components, and have provided important constraints upon theoretical models for interstellar shock waves (Neufeld et al.\ 2006a, 2007).

While warm molecular hydrogen has been studied primarily through spectroscopic observations with the IRS on {\it Spitzer}, photometric observations with the Infrared Array Camera (IRAC) are also possible.  As noted recently by Reach et al.\ (2006; see their Fig.\ 1), the $v=1-0$ O(5) and $v=0-0$ S(4) - S(13) transitions of H$_2$ fall within the IRAC bandpass and can contribute significantly to the intensities observed with IRAC; in some sources, the IRAC intensities are potentially dominated by H$_2$ line emissions.  In this paper, we investigate further the utility of IRAC observations in mapping the distribution of warm molecular hydrogen.  In \S 2, we discuss IRAC and IRS observations carried out toward the supernova remnant IC443.  In \S 3, we compare the spatial distribution of the H$_2$ line emissions, observed with IRS, with the photometric maps obtained with IRAC.  In \S 4, we discuss how the relative intensities within the four IRAC bands can be used to 
constrain the physical conditions within the warm molecular gas. 
  
\section{Observations}

In a previous paper (Neufeld et al.\ 2007), we have discussed spectroscopic mapping observations of warm molecular hydrogen, carried by {\it Spitzer}/IRS toward small ($1^\prime \times 1^\prime$) regions in the SNR W44, W28, 3C391 and IC433.  Full details of the observations and data reduction procedure have been described by Neufeld et al.\ (2006a; 2007) and will not be repeated here.  In the case of IC443, our IRS observations targeted a region containing clump C (in the nomenclature of Denoyer 1979), and -- of the four SNR sources observed -- revealed the spectrum most strongly dominated by line emissions (Neufeld et al.\ 2007, see their Fig.\ 8).  In Figure 1, we reproduce the observed IRS spectrum for a 25$^{\prime\prime}$ diameter circular region centered on clump C (black curve), superposed on the spectral response functions for the four IRAC bands (Fazio et al.\ 2004; blue).  The red curve represents a model H$_2$ spectrum for spectral regions shortward of the IRS wavelength range; it will be discussed in \S 4 below.  Figure 1 suggests that the Band 4 (8~$\mu$m) bandpass \ree{could be} dominated by H$_2$ S(4) and S(5) emissions, and the Band 3 (5.6~$\mu$m) bandpass by H$_2$ S(6) and S(7).  Emissions from polycyclic aromatic hydrocarbons (PAHs) do not appear to contribute significantly to the Band 3 and Band 4 intensities in IC433 (in contrast to some of the other supernova remnants observed by Neufeld et al.\ 2007.)

To evaluate IRAC observations as a potential probe of warm molecular hydrogen, we extracted IRAC maps of IC443 from the {\it Spitzer} archive.  Two IRAC fields within IC433 are available in the archive, both observed in Program 68 and one of them encompassing the region mapped with IRS.  The location of the \re{regions} observed with {\it Spitzer} IRAC and IRS are shown in Figure 2, superposed on a 2MASS map of the the entire remnant (Rho et al.\ 2001).  The southernmost field observed with IRAC has an angular area that is $\sim 70$ times that of the IRS field it encompasses.  We made use of maps generated by the SSC pipeline mosaicker, subtracting the median intensity from each band to obtain the Band 1 -- 4 maps shown in Figure 3.  Our use of median-subtraction provides a rough removal of any diffuse sky background; the median values that we subtracted were 0.23, 0.29, 1.93 and 8.10 MJy~sr$^{-1}$ respectively for Bands 1, 2, 3, and 4.  \ree{As discussed in \S3 below, this diffuse component may show variations on spatial scales smaller than those of the entire IRAC map.  While the overall accuracy of the IRAC flux calibration is believed to be better than 3$\%$ for point sources (Reach et al.\ 1995, based upon the assumption that models for A stars are accurate to much better than $3\%$), the extended source calibration is less certain.  For sources (galaxies) of angular diameter, $\theta$, greater than $\sim 9^{\prime \prime}$, an extended source aperture correction is found to be necessary: the measured fluxes must be multiplied by a correction factor that decreases monotonically from 1.0 at $\theta = 9^{\prime \prime}$ to values at $\theta > 300^{\prime \prime}$ of 0.91, 0.94, 0.71, and 0.74 respectively for Bands 1, 2, 3, and 4.  This correction factor is needed because the point spread functions for Bands 3 and 4 show more extended wings than expected and because there is a extremely diffuse scattering that distributes a portion of the incident flux over the entire detector array (Reach et al.\ 2005.) In the context of the present observations, the appropriate correction factor is ill-defined because the maps show structure on a range of size scales.  However, since our primary interest is in the brightest emission knots that are generally less than $10^{\prime \prime}$ in extent, we will adopt the point source calibration without correction.  The implications of the flux calibration uncertainties will be discussed further in \S 4 below.}

\section{Comparison of the IRS and IRAC maps}

Given the \re{spectral} response functions (Fazio et al.\ 2004) plotted in Figure 1, \re{H$_2$ emission alone is expected to yield the following IRAC intensities}:

$${I_{\rm Band \,3}({\rm H}_2) \over {\rm MJy\, sr}^{-1}} =  {0.909\, I[{\rm H}_2\,S(6)] + 0.700 \, I[{\rm H}_2\,S(7)] \over {\rm 10^{-4}   \, erg \, cm^{-2} \, s^{-1} \, sr^{-1}}} \eqno(1)$$

$${I_{\rm Band \,4}({\rm H}_2) \over {\rm MJy\, sr}^{-1}} =  {0.783\, I[{\rm H}_2\,S(4)] + 0.566 \, I[{\rm H}_2\,S(5)] \over {\rm 10^{-4}   \, erg \, cm^{-2} \, s^{-1} \, sr^{-1}}} \eqno(2)$$
\vskip 0.1 true in
We have rebinned the IRS and IRAC data on a common grid and compared, pixel-by-pixel, the IRAC-measured intensities, $I_{\rm Band \,3}({\rm IRAC})$ and $I_{\rm Band \,4}({\rm IRAC})$, with the quantities  $I_{\rm Band \,3}({\rm H_2})$ and $I_{\rm Band \,4}({\rm H_2})$ given by equations (1) and (2) above.\footnote{This comparison has revealed modest astrometry errors in the output of our IRS mapping software (Neufeld et al.\ 2006a; the errors are smaller than 1$^{\prime\prime}$ at the center of the IRS-mapped field, rising to up 4$^{\prime\prime}$ at the edges).  A small shift ($\le \re{1} ^{\prime\prime}$) of the central position in the IRS maps, combined with a 12$\%$ stretch along the slit direction, bring the IRS and IRAC maps into excellent agreement.}  Figure 4 uses scatter plots to present the results, with the best linear fit superposed.  Most points lie very close to the best linear fit.  Small positive intercepts (2.0 and 2.5 MJy~sr$^{-1}$ respectively for Bands 3 and 4) on the vertical axis for the best-fit line may suggest that the diffuse background in the IRS-mapped region is somewhat larger than that for the IRAC-mapped region as a whole.  
\ree{Alternatively, or in addition, they may result from instrumental scattering of radiation from the swath of bright emission within the IRS-mapped region (see discussion at the end of \S 2).  The fact that the ratio of intercepts (2.0/2.5 = 0.8) is comparable to the typical Band 3 / Band 4 line ratio (0.7) lends support to the second possibility.  In any case, the intercepts are small ($\sim 15\%$) relative to the midpoints of the range of measured fluxes, and we shall henceforth assume Bands 3 and 4 to be largely dominated by H$_2$.}
The slopes on the best-fit linear relationships are 1.10 and 1.03 respectively for Bands 3 and 4.  Given the likely uncertainties in the flux calibration for the IRS line maps -- estimated by Neufeld et al.\ (2007) as $\simlt 30 \%$ -- these values are remarkably close to unity. 

In Figure 5, we compare maps of IC433 obtained from IRS observations using equations (1) and (2) with those obtained using IRAC.  If we correct the former by multiplicative factors of 1.16 and 1.12 respectively for Bands 3 and 4, we obtain a remarkable accord between the H$_2$ maps derived using the two different instruments.  The only significant difference is apparent at offset $(-12^{\prime\prime}, -43^{\prime\prime})$, where a point continuum source shows up in the IRAC maps but not the H$_2$ spectral line maps.  This is the $K=10.4$ mag 2MASS source 06174344+2221060.
The comparisons presented in Figures 4 and 5 indicate that IRAC is a powerful tool for studying warm molecular hydrogen in those sources where the mid-IR spectrum is dominated by H$_2$ line emission.  
While mapping a $\sim 1$ square arcminute region with IRS required an observing time of 93 minutes, the IRAC map of area $\sim 70$ square arcminutes required only 13 minutes to observe.  In the next section, we will consider how the IRAC colors (i.e.\ ratios of intensities for different bands) can be used to constrain the physical conditions in the warm molecular gas.

\section{Physical conditions in the gas traced by IRAC}

In Figures 6 -- 8 (lower panels), we present maps of the ratio of \ree{three independent pairs} of IRAC band intensities.  In these figures, we suppressed regions in which the IRAC intensities are small, with a median-subtracted Band 3 intensity less than 5~MJy~sr$^{-1}$, because here the noise becomes significant    
\ree{and the contribution of diffuse emission may be important.  This threshold value is a factor 2.5 as large as the positive intercept obtained in Figure 4 (see \S3); we have determined that the adoption of a lower threshold of 3~MJy~sr$^{-1}$ has little effect upon the distribution of observed band ratios.}

The upper panels show histograms (arbitrary scale) that indicate the relative number of pixels for which the logarithm of the band ratio lies in a given range.  The colors appearing underneath the histogram curve are matched to those appearing in the lower panel, so the histogram serves as a color bar for the band ratio maps.  Vertical dashed lines indicate the values of log(band ratio) corresponding to the lower sextile, median, and upper sextile of the distribution.  Figures 6 -- 8 show that different IRAC bands exhibit different degrees of correlation.  Bands 1 and 2 are most closely correlated (narrowest histogram for the band ratio), while Bands 1 and 4 are the least so.  All 3 histograms indicate a skew distribution for the band ratios, with an excess of pixels for which the shorter-wavelength of the two bands is abnormally large.  A cursory inspection of the ratio maps indicates that the right-hand tail of the distribution (colored yellow) is mainly accounted for by discrete point sources, i.e.\ stars.

In interpreting these band ratios, we have determined which H$_2$ lines fall in each IRAC band.  In Table 1, we list the contribution of various H$_2$ lines to each IRAC band, together with the upper state energies.  The former, computed using the \re{spectral response} functions plotted in Figure 1, are in units of $10^4 \rm \,  MJy\,sr^{-1}/(erg \, cm^{-2} \, s^{-1} \,sr^{-1})$.  For example, the entry 0.909 appearing opposite H$_2$ S(6) indicates that an H$_2$ S(6) line intensity of $10^{-4} \, \rm erg \, cm^{-2} \, s^{-1} \,sr^{-1}$ will add a contribution of $0.909 \rm \, MJy\,sr^{-1}$ to the intensity measured in IRAC Band 3 (in accord with eqn.\ 1).  Assuming that the H$_2$ level populations are controlled by collisional excitation by H$_2$ and He, we then used a statistical equilibrium calculation to determine the expected IRAC band ratios as a function of the gas temperature and pressure.  Here we adopted the collisional excitation rates of Flower \& Roueff \re{(1998a, 1998b, 1999) and Flower, Roueff \& Zeippen (1998)}.  We also assumed the H$_2$ ortho-to-para ratio to have attained its equilibrium value of 3, consistent with the results of IRS spectroscopy for the S(4) -- S(7) transitions (Neufeld et al.\ 2007), \re{and adopted a $n({\rm He})/n({\rm H}_2)$ ratio of 0.2}

In Figure 9, we show the results of our statistical equilibrium calculation as a function of temperature and density (blue curves).  Here, the expected IRAC band ratios are shown in a two-color diagram, with the logarithm of the
Band-2-to-Band-4 ratio plotted on the horizontal axis and that of the Band-3-to-Band-4 ratio on the vertical axis.  Both ratios are strongly increasing functions of temperature, as expected because the Band 4, Band 3, and Band 2 wavelength ranges contain H$_2$ lines of successively higher excitation (see Table 1).
The open triangle represents the median band ratios observed in the IRAC maps, with the error bars designating the upper and lower sextiles.  \ree{While the open triangle with the solid error bars is obtained with the standard point source flux calibration that we consider appropriate for this application (see discussion in \S 2 above), the open diamond with the dotted error bars is obtained for flux calibration appropriate to a highly extended source.  Clearly, regardless of the exact procedure adopted to obtain the IRAC flux calibration,} no single combination of temperature and density comes close to yielding the observed band ratios. 

This key result implies either (1) that a {\it mixture} of gas temperatures is present along every sight-line; and/or (2) that Band 2 is dominated by emission from some source other than H$_2$.  The first of these is {\it known} to be true, based upon IRS observations of the H$_2$ S(0) -- S(7) rotational emissions.  For IC443 and every other source observed by Neufeld et al.\ (2006a) and (2007), the H$_2$ rotational diagrams exhibit a distinct positive curvature that indicates multiple temperature components to be present.  In evaluating the second possibility, we have considered possible contributions from the Br$\alpha$ recombination line of atomic hydrogen (at 4.05~$\mu$m) and the $v=1-0$ band of carbon monoxide (at 4.6~$\mu$m).  The first of these can be ruled out from observed upper limits on the Br$\gamma$ line (Burton et al.\ 1988), and the latter \ree{by ground-based observations of the 4.687 -- 4.718~$\mu$m region (Richter et al.\ 1995; this region is inaccessible to {\it Spitzer}/IRS).  Although the possible contribution of CO vibrational emissions to the IRAC Band 2 intensity has been discussed previously for shocked molecular gas (e.g. Reach et al.\ 2006), we find that the negligible contribution observed in IC443C is entirely unsurprising; this conclusion is discussed further at the end of \S 4, following our discussion of the gas temperature distribution.}

In an attempt to model an admixture of gas temperatures, we have adopted a simple power-law distribution of the form $$dN = aT^{-b} dT, \eqno(3)$$
where $dN$ is column density of gas at temperature between $T$ and $T+dT$; and $a$ and $b$ are parameters that we will adjust to fit the band intensities.  Here, we adopt a lower limit, $T_{\rm min}$ on the gas temperature of 300~K and an upper limit, $T_{\rm max}$, of 4000~K.  The exact lower limit is unimportant, since gas cooler than a few hundred K is too cold to emit significant H$_2$ emission in the IRAC wavelength range; the upper limit represents the temperature above which H$_2$ is rapidly dissociated (Le Bourlot et al.\ 2002).  The coefficient $a$ in equation (3) can then be written
$$a= {N(>300\,{\rm K})(b-1) \over (T_{\rm min}^{1-b}-T_{\rm max}^{1-b})}, $$ 
where $N(>300\,{\rm K})$ is total column density of molecular hydrogen warmer than 300~K.
\re{In comparing the predictions of our excitation model with the observations, we adopt an assumed extinction $A_V = 13.5$~mag, based upon the estimate of Richter et al. (1995) for
the 2.1 $\mu$m extinction, together with the interstellar extinction curves of Weingartner \& Draine (2001; ``Milky Way, $R_V=3.1$").} Red curves in Figure 9 show the expected band ratios for an {\it admixture} of gas temperatures, with the effects of extinction included.  For a suitable choice of the assumed power-law index, $b$, and gas density, $n({\rm H}_2)$, the observed band ratios are now consistent with the H$_2$ excitation model.  The median Band 3 / Band 4 and Band 2 / Band 4 intensity ratios are best fit by \re{$b \sim 4.5$ and $n({\rm H}_2) \sim 10^6 \rm \, cm^{-3}$.  Unfortunately, because the observed band ratios lie close to the locus obtained in the high-density limit (i.e.\ in local thermodynamic equilibrium), the inferred density is strongly dependent upon the measured band ratios and is therefore quite uncertain.} 

The results of our excitation calculations are also plotted in Figures 6 -- 8 (upper panel), for the second case where an admixture of temperatures is assumed.  In each figure, the expected band ratio -- with the effects of extinction included -- is plotted {\it on the horizontal axis} as a function of the power-law index, $b$ (plotted on the vertical axis), and for H$_2$ densities of 10$^4$, 10$^5$, 10$^6$ and $10^7 \rm \, cm^{-3}$.  One consistent behavior is particularly noteworthy: those ratios \ree{with the strongest dependence on $b$} show the largest variations within the map.  Thus the Band-1-to-Band-2 ratio, which shows the smallest spatial variation, has the weakest dependence upon $b$.  This strongly suggests that the \ree{spatial} variations in the measured band ratios reflect real variations in the temperature distribution of the emitting gas.

In Figure 10, we present additional color-color diagrams for a selection of three different pairs of band ratios.  Instead of simply showing the median and upper and lower sextile for the distribution of observed band ratios as we did in Figure 9, we now show the two-dimensional distribution function.  Thus, the false color image (coded from black [smallest value] to blue to white [largest value]) represents the number of pixels for which the band ratios show a given pair of values.  The white regions in these maps (representing the most probable combination of line ratios) are clearly elongated parallel to the lines of constant density. Again, this behavior suggests that the variations in the measured band ratios reflect real variations in the temperature distribution of the emitting gas.  

Figure 10 reveals a significant discrepancy between the H$_2$ excitation model and the observations.  While all the plotted band ratios suggest a power-law index $b$ varying over the range 3 -- 6, the best-fit density is not consistently determined.  In particular, the relative strengths of Bands 2 -- 4 suggest a typical density $n({\rm H}_2) \sim 10^6 \rm \, cm^{-3}$ (with large error bars on the exact value), whereas the relative strength of Band 1 argues for a higher density $\sim 10^7 \rm \, cm^{-3}$.  Both values are considerably larger than the density $\sim \rm few \times 10^4  \rm \, cm^{-3}$ inferred previously from an analysis of the HD R(4)/R(3) line ratio (Neufeld et al.\ 2006b).

\re{These discrepancies may result from the presence of a mixture of gas densities within the source.  They may also reflect a shortcoming in our treatment of the H$_2$ excitation, in that we only include the effects of collisions with H$_2$ and He.  For the $v = 1-0$ transitions that contribute to Band 1, our treatment may significantly underestimate the line strengths, because we neglect the effects of collisional excitation by atomic hydrogen.  For vibrational transitions of H$_2$ -- unlike pure rotational transitions in which excitation by H is {\it less efficient}-- the cross-section for excitation by H can exceed that by H$_2$ by several orders of magnitude (see Le Bourlot et al.\ 1999, their Table 3, for a tabulation of the relevant critical densities for which collisional excitation by H or H$_2$ dominates spontaneous radiative decay.)  At the typical temperatures indicated by our observations of IC443, we find that the excitation of vibrational emissions by collisions with H dominates for $n({\rm H})/n({\rm H}_2)$ ratios greater than $\sim 0.025$;  thus, collisional excitation by atomic hydrogen can be important even in gas that is primarily molecular.  In shocked molecular clouds, atomic hydrogen can be produced by collisional dissociation of H$_2$, and by reaction of H$_2$ with O and OH.  Theoretical models for shock waves (e.g.\ Wilgenbus et al.\ 2000) suggest that significant atomic hydrogen abundances can be achieved behind shocks that are fast enough to produce H$_2$ at temperatures of a few thousand K.  For H abundances of this magnitude, the excitation of H$_2$ vibrational transitions can be dominated by collisions with H. } 

\re{To assess the typical contribution of various spectral lines to the IRAC band intensities, we have considered a ``typical excitation model" for the excitation of H$_2$ in IC443.
Here we adopt $b \sim 4.5$ and $n({\rm H}_2) = 10^6 \rm \, cm^{-3}$ to obtain the pure rotational line ratios.  For rovibrational transitions, we adopt a larger H$_2$ density, $n({\rm H(Richter et al.\ 1995)}_2) \sim 10^7 \rm \, cm^{-3}$, as an approximate method of treating of the effects of excitation by atomic hydrogen and/or the presence of multiple density components. For this typical excitation model, we have computed the H$_2$ line spectrum expected in the $3 - 5$~$\mu$m region covered by IRAC Bands 1 and 2.  This spectrum is shown in Figure 1 (red curve, for an assumed spectral resolution $\lambda/\Delta\lambda=60$ FWHM) , and the fractional contribution of each line to each IRAC band is given in Table 1 (rightmost column).  For this excitation model, Bands 1, 2, 3, and 4 are respectively dominated by H$_2$ $v = 1-0$ O(5), $v = 0-0$ S(9), $v = 0-0$ S(7), and $v = 0-0$ S(5).}

\ree{Although a significant contribution of CO vibrational emissions to IRAC Band 2 is ruled out by direct observation in IC443C (Richter et al.\ 1995, discussed above), we have also considered the conditions under which CO v=1-0 emissions might be important in other sources.}  Here we made use of recent theoretical calculations of the rate coefficients for excitation of CO vibrational transitions by He (Cecchi-Pestellini et al.\ 2002) and H (Balakrishnan et al.\ 2002).  For excitation by H$_2$, we adopted the expression given Thompson (1973, his equations [7] and [8], with his parameter $A$ taken as 68, in accord with the laboratory measurements of Millikan \& White 1963.)  In Figure 11, we plot the fractional contribution of CO emissions to IRAC Band 2, as a function of the H$_2$ density and for an assumed CO/H$_2$ abundance ratio of $10^{-4}$.  As with the vibrational excitation of H$_2$, the cross sections for excitation by H greatly exceed those for excitation by He or H$_2$, so the results depend upon the H/H$_2$ abundance ratio.  Results are given for H/H$_2$ abundance ratios of 0, 0.01, 0.1, and 1 (from bottom to top), and for temperature power-law indices, $b$, of 4 (blue curves) and 5 (red curves).  As before, an He/H$_2$ ratio of 0.2 is assumed.  For H/H$_2$ ratios $\le 10^{-2}$, CO vibrational emissions contribute significantly (i.e. above the 10$\%$ level) only at densities $\ge 3 \times 10^{7} \rm \, cm^{-3}$. At a density of $10^6 \rm cm^{-3}$, the contribution from CO is only $\simlt 1\%$ if the H/H$_2$ ratio $\le 10^{-2}$.  Even for an H/H$_2$ ratio of unity, the CO contribution is $\simlt 20\%$ at $n({\rm H}_2) = 10^6\, \rm cm^{-3}$; while shock models do allow for the possibility of large H/H$_2$ ratios near the threshold for dissociation, we consider it implausible that the effective H/H$_2$ ratio for the excitation of CO vibrational emissions could be as large as unity, because most of the CO emission originates in relatively cool gas that has been subject to shocks of velocity much less than the dissociation threshold.

\section{Discussion}

The excitation of molecular hydrogen behind interstellar shock waves has been the subject of considerable discussion over the past two decades.  In the late 1980's, with the advent of spectral line mapping at near-IR wavelengths, two key features of the H$_2$ emission were noted: (1) the relative line strengths observed toward shocked regions are \ree{typically} inconsistent with a single excitation temperature (Burton et al.\ 1989); and (2) even line ratios that are strongly dependent upon the gas temperature often show a remarkable constancy throughout a given source (Brand et al.\ 1989).  \ree{These behaviors have subsequently been observed in a wide variety of sources, and are particularly evident in {\it Spitzer} observations where the S(1) -- S(7) pure rotational lines are mapped over a common region
(e.g.\ in both the Herbig Haro objects -- HH54 and HH7 -- observed by Neufeld et al.\ 2006a; and in all the supernova remnants -- IC443, W28, W44, 3C391 -- observed by Neufeld et al.\ 2007).}

It was recognized (Smith \& Brand 1990) that these behaviors are unexpected for single planar shocks that are of ``C-type'' in the designation of Draine (1980).  In such shocks, ion-neutral drift is a critical feature of the physics (Mullan 1971), allowing the velocities of the ionized and neutral species to change continuously (hence the designation ``C-type'' for ``continuous'') in a warm region of roughly uniform temperature: as a result (a) the rotational diagrams for the H$_2$ level populations behind such shocks show little curvature (e.g. Neufeld et al.\ 2006a, Appendix B); and (b) temperature behind the shock front is a strong function of the shock velocity\footnote{ While planar C-type shocks were subsequently shown to be unstable (Wardle 1991), numerical simulations of the non-linear development of the instability (Neufeld \& Stone 1997; MacLow \& Smith 1997) suggested that both results (a) and (b) still apply.}. 
The former is inconsistent with observational feature (1) above, and the latter is inconsistent with (2) unless the shock velocity is remarkably constant throughout the mapped region.  

Two separate explanations were offered to account for the observations, i.e.\ the key features (1) and (2) described above.  First, Burton et al.\ (1989) noted that the cooling region behind a ``J-type'' shock might successfully explain the H$_2$ line ratios.  In a ``J-type'' shock, where the ion and neutral species are well-coupled and show velocities that jump discontinuously at the shock front (hence the designation ``J-type'' for ``jump''), the gas is initially heated to a high temperature and then cools radiatively, resulting in an admixture of gas temperatures.  Provided the shock is not fast enough to destroy molecular hydrogen, J-type shocks yield a rotational diagram with significant curvature but which shows a relatively weak dependence upon the shock velocity.  

As a second and alternative explanation, Smith, Brand \& Moorhouse (1991; hereafter SBM) suggested bow shocks as the source of the observed H$_2$ emissions.  Such shocks result when a supersonic flow encounters a clump of material or when a supersonic clump travels through an ambient medium.  In SBM's model, multiple unresolved bow shocks are assumed to be present in each telescope beam.  At the head of each bow shock, where the gas moves perpendicular to the shock front, the effective shock velocity is greatest; here the shock may become dissociative and of J-type.  On the sides of the bow shock, where the material enters the shock front at an oblique angle, the shock velocity is smaller and the shock is non-dissociative and of C-type.  Assuming a parabolic shape for the shock surface, and decomposing a bow shock into a set of planar shocks of varying shock velocity, SBM considered the resultant H$_2$ emission. 
The H$_2$ rotational diagram is indeed expected to show significant curvature, consistent with the observations.  Furthermore, provided that the shock velocity at the head of the bow shock is sufficient to dissociate molecules, SBM showed that the H$_2$ line ratios are independent of the flow velocity.  

Bow shocks offer several advantages over J-type shocks in explaining the observations.  First, given the typical ionization level and magnetic field strengths in molecular clouds, non-dissociative shock waves are expected to be of C-type, not J-type.  Second, high-resolution observations of shocked H$_2$ (e.g. Schultz et al.\ 1999), carried out after the explanations of Burton et al.\ (1989) and SBM were offered, have indeed revealed the presence of bow shocks that were unresolved in previous observations.  Finally, recent observations of the S(0) -- S(7) rotational transitions of H$_2$ (Neufeld et al.\ 2006a), carried out with {\it Spitzer}/IRS toward six separate sources, have shown that the 
H$_2$ ortho-to-para ratio is smaller than its equilibrium value, {\it with the cooler components along every sight-line showing the greatest departures from equilibrium.}  This behavior is simply inconsistent with the J-type shock model, in which the warmer gas components evolve to become the cooler components.  To the contrary, the cooler components must arise in separate shocked regions. 

We can compare the power-law index, $b$, derived in \S 4 above with the predictions of bow shock models.  For a curved shock front with a paraboloidal shape, material crossing the shock front with a perpendicular velocity between $v$ to $v+dv$ is associated with a shock surface area $dA \propto v^{-4} dv$ (Smith \& Brand 1990); this proportionality applies for all velocities less than the maximum shock velocity, $v_{\rm max}$, attained at the head of the bow shock.  Based upon a fit to the planar C-type shock models of Kaufman \& Neufeld (1996), Neufeld et al.\ (2006a, Appendix B) found that the warm shocked material could be adequately approximated by a slab of constant temperature $\propto v^{1.35}$ and column density $N ({\rm H}_2) \propto v^{-0.75}$.   These proportionalities yield a power-law temperature distribution, with the mass at temperature between $T$ and $T+dT$ obeying
$$dM \propto N({\rm H}_2)  dA \propto v^{-4.75} dv \propto T^{-(3.75+1.35)/1.35} dT \propto T^{-3.78} dT 
\eqno(3).$$

For bow shocks with a maximum shock velocity sufficient to dissociate H$_2$ at the head of the bow, this leads to a 
power-law index $b \sim 3.8$ extending to the maximum possible H$_2$ temperature, $T_{max}$.  For slower bow shocks, the temperature distribution extends to some temperature lower than $T_{max}$.  Thus an admixture of bow shocks, with a range of velocities some of which insufficient to achieve temperature $T_{\rm max}$ at the head of the bow shock, could lead to an overall temperature distribution which is steeper than $dN \propto T^{-3.8} dT$ (i.e.\ to values of $b$ greater than 3.8).

These simple predictions are broadly consistent with the observations, which indicate values of
$b < 3.8$ to be quite rare.  In this scenario, the observed spatial variations in $b$ could result from variations in the distribution of velocities at the heads of a set of unresolved bow shocks.  We note, however, that our assumption of parabolic bow shocks in steady-state is a highly idealized one, and that numerical simulations would be needed to provide a more realistic prediction for the shock structure and resultant temperature distribution in the presence of possible instabilities (including the Wardle instability).

\section{Summary}

\noindent 1.  IRAC maps of IC443C, extracted from the {\it Spitzer} archive, are strikingly similar to spectral line maps of the H$_2$ pure rotational transitions, obtained with the IRS instrument.  

\noindent 2.  IRS spectroscopy indicates that IRAC Bands 3 and 4 are dominated in this source by the H$_2$ $v=0$--0 S(5) and S(7) transitions, respectively.  Modeling of the H$_2$ excitation suggests that Bands 1 and 2 are dominated by H$_2$ $v = 1-0$ O(5) and $v=0$--0 S(9). 

\noindent 3.  Mapping H$_2$ with IRAC presents several advantages and disadvantages relative to mapping with IRS.  IRAC mapping:

(a) is a factor $\sim 500$ faster in area mapped per unit observing time;

(b) provides access to additional H$_2$ transitions: $v=1-0$ O(5), which is sensitive to collisional excitation by atomic hydrogen, and $v=0-0$ S(9), which is of higher excitation than any transition accessible to IRS;

(c) fails to probe lower excitation transitions -- S(0), S(1), S(2), and S(3) -- that {\it are} accessible to IRS;

(d) fails to probe the H$_2$ ortho-to-para ratio, since each band is sensitive to both ortho- and para-H$_2$ transitions;

(e) is only possible in sources with H$_2$-dominated mid-IR spectra.

\noindent 4.  Large maps of the H$_2$ emission in IC433, obtained with IRAC, show band ratios that are inconsistent with a single temperature component.  The relative strengths of IRAC Bands 2, 3, and 4 are consistent with pure H$_2$ emission from shocked material with a power-law distribution of gas temperatures.  CO vibrational emissions \ree{do not}  contribute significantly to the observed Band 2 intensity.

\noindent 5.  The relative strength of Band 1 is larger than that predicted in models for the collisional excitation of H$_2$ by H$_2$ and He.  The observations can be reconciled with a model in which a large atomic hydrogen abundance enhances the H$_2$ vibrational emissions believed to dominate IRAC Band 1.

\noindent 6.   Assuming that the column density of H$_2$ at temperatures $T$ to $T+dT$ is proportional to $T^{-b}$ for temperatures up to 4000~K, we obtained a typical estimate of 4.5 for $b$.  The power-law index, $b$, shows variations over the range $\sim 3 - 6$ within the set of different sight-lines probed by the maps, with the majority of sight-lines showing $b$ in the range $4-5$.

\noindent 7.  The observed power-law index is consistent with the predictions of simple models for paraboloidal bow shocks.

\acknowledgments

This work, which was supported in part by RSA agreement 1263841, is based on observations made with the {\it Spitzer Space Telescope}, which is operated by the Jet Propulsion Laboratory, California Institute of Technology, under a NASA contract.
We gratefully acknowledge the additional support of grant NAG5-13114 from NASA's Long Term Space Astrophysics (LTSA) Research Program.  \re{This research has made use of the BASECOL database (http://amdpo.obspm.fr/basecol) -- devoted to the rovibrational excitation of molecules -- and of NASA's Astrophysics Data System.}

\clearpage

\begin{deluxetable}{lccccr}
\tablewidth{0pt}
\tablecaption{Contribution of H$_2$ line emission to the IRAC bands} 
\tablehead{
Transition & Wavelength & Upper state energy & IRAC & Contribution$^a$ & Fractional \,\,\, \\
           & ($\mu$m)   & $E_U/k$ in Kelvin  &           &                  & contribution$^b$}
\startdata
H$_2$ $v=1-0$ \,\,\, O(5) &       3.2350  & 6952 &  1  &     0.408  &    44 \\
H$_2$ $v=1-0$ \,\,\, O(6) &       3.5008  & 7584 &  1  &     0.572  &    14 \\
H$_2$ $v=1-0$ \,\,\, O(7) &       3.8074  & 8366 &  1  &     0.637  &    29 \\
H$_2$ $v=0-0$ \,\,\, S(13)&       3.8472  & 17445&  1  &     0.613  &    13 \\ 
H$_2$ $v=0-0$ \,\,\, S(12)&       3.9969  & 15542&  2  &     0.233  &    2 \\
H$_2$ $v=0-0$ \,\,\, S(11)&       4.1815  & 13704&  2  &     0.610  &    22 \\ 
H$_2$ $v=0-0$ \,\,\, S(10)&       4.4100  & 11941&  2  &     0.744  &    14\\
CO $v=1-0$ \,\,\,  &              4.6     & 3080 &  2  &   0.696  &   see \S4 \\
H$_2$ $v=0-0$ \,\,\, S(9) &       4.6947  & 10262&  2  &     0.721  &    62 \\
H$_2$ $v=0-0$ \,\,\, S(8) &       5.0531  & 8678 &  3  &     0.137  &     3 \\  
H$_2$ $v=0-0$ \,\,\, S(7) &       5.5112  & 7197 &  3  &     0.700  &    63 \\
H$_2$ $v=0-0$ \,\,\, S(6) &       6.1086  & 5830 &  3  &     0.909  &    34\\
H$_2$ $v=0-0$ \,\,\, S(5) &       6.9095  & 4587 &  4  &     0.566  &    70 \\
H$_2$ $v=0-0$ \,\,\, S(4) &       8.0251  & 3475 &  4  &     0.783  &    30 \\
\enddata
\tablenotetext{a}{in units of $10^4 \rm \,  MJy\,sr^{-1}/(erg \, cm^{-2} \, s^{-1} \,sr^{-1})$}
\tablenotetext{b}{Fractional contribution to the band, in percent, for our typical excitation model (see \S4).}
\end{deluxetable}

\clearpage

\begin{figure}
\includegraphics[scale=0.65,angle=0]{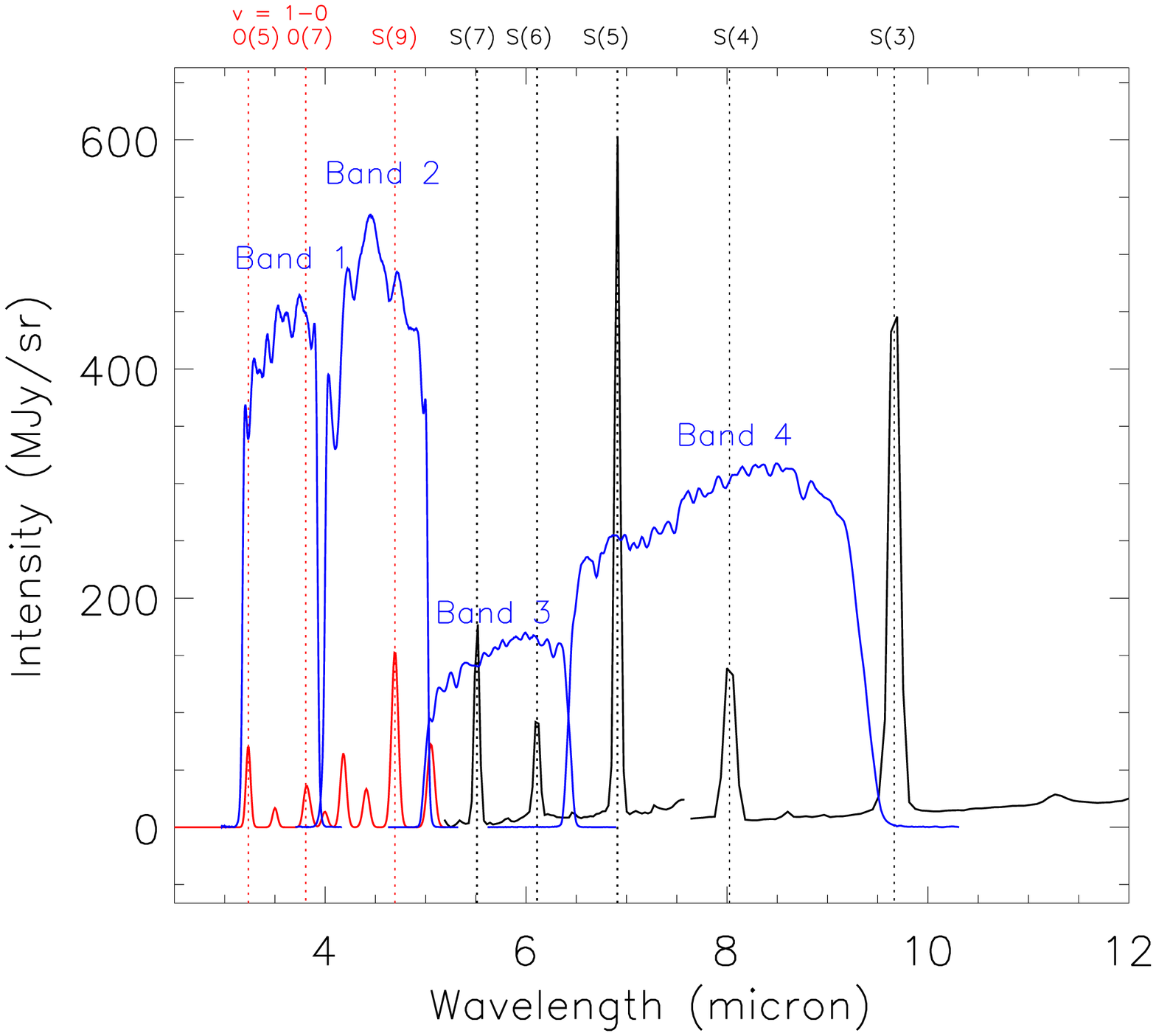}

\noindent{Fig.\ 1 -- Black curve:
average spectrum observed by IRS for a $25^{\prime\prime}$ 
(HPBW) diameter circular aperture centered at  
$\alpha=06^{\rm h}$ 17$^{\rm m}$ 42.$^{\rm \!\! s}$5, 
$\delta=+22^{\rm o}$ 21\amin\ 29\asec\ (J2000)
in IC443C (corresponding to offset $(-24^{\prime\prime},-20^{\prime\prime})$ in Figure 3 below).  
The spectra shown here are not background-subtracted, since no off-source measurements were made, and thus the continuum flux level must be regarded as somewhat uncertain.
Blue curves: relative response function for the four IRAC photometric bands (Fazio et al.\ 2004; shown here in arbitrary units).
Red spectrum: predicted H$_2$ spectrum for our typical excitation model  
(described in \S4 below).}

\end{figure}
\clearpage

\begin{figure}
\includegraphics[scale=0.60,angle=0]{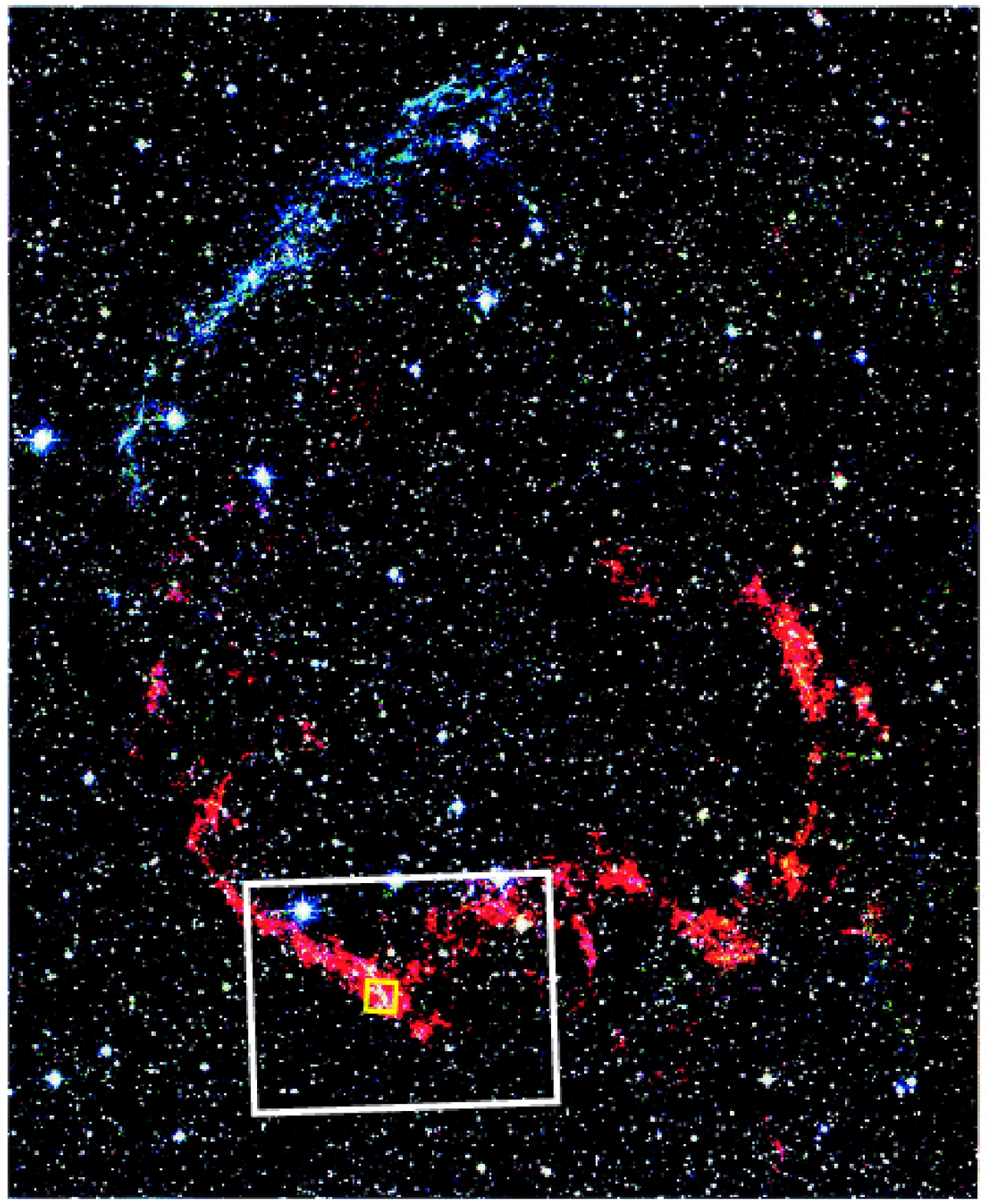}

\noindent{Fig.\ 2 -- Finder chart for IC443.  The regions mapped by IRAC (white rectangle which shows the region common to all four bands) and IRS (yellow square) are superposed on a 2MASS image of IC443 (from Rho et al.\ 2001).  Here, the blue, green and red channels represent the J (1.25$\mu$m), H (1.65$\mu$m), and K$_S$ (2.17$\mu$m) intensities, the latter being dominated in the southern ridge by $v=1-0$ emissions from H$_2$ (Rho et al.\ 2001).}

\end{figure}
\clearpage

\begin{figure}
\includegraphics[scale=0.80,angle=0]{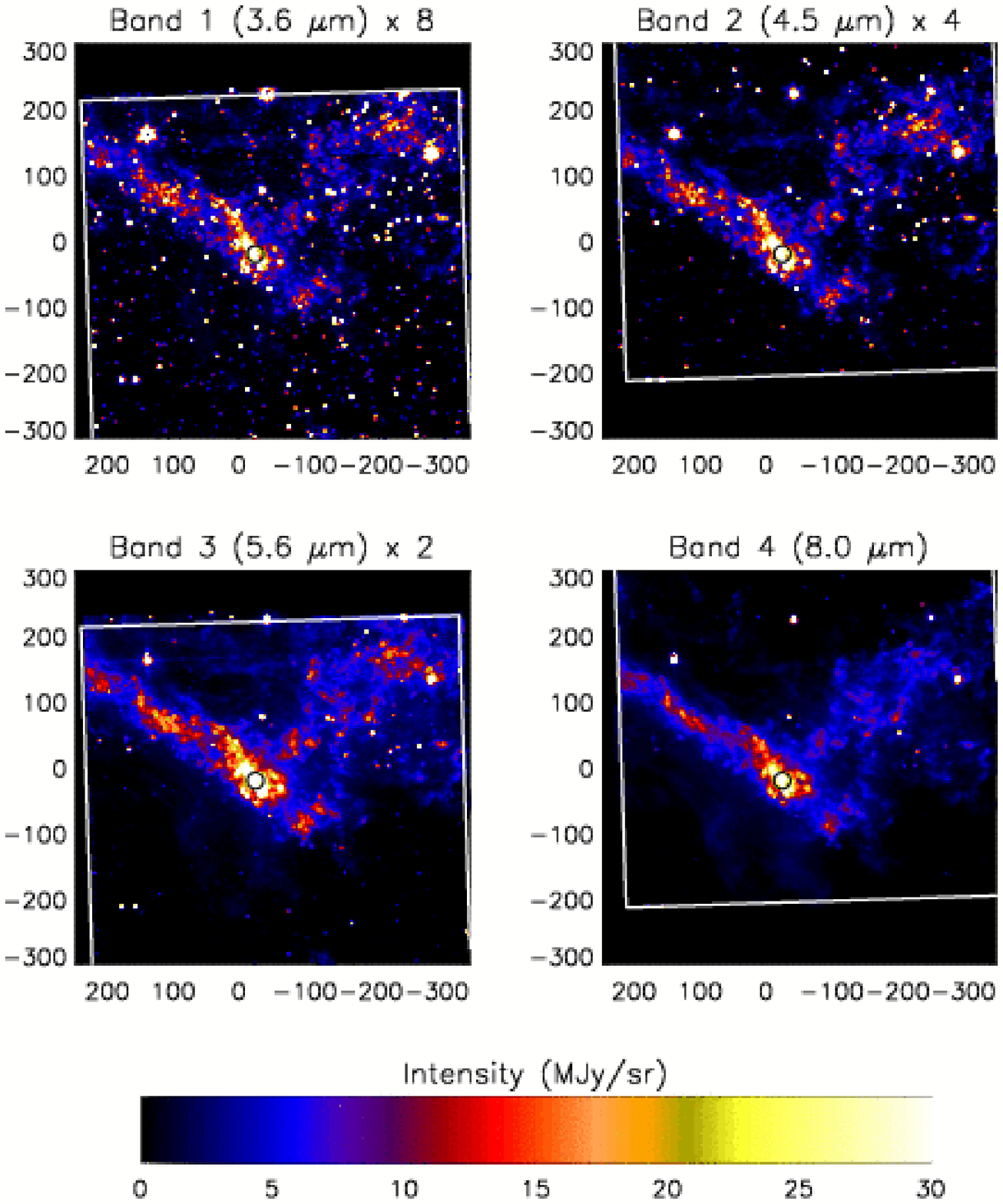}

\noindent{Fig.\ 3 -- Intensities observed in the four IRAC bands 
toward IC443C.  The
horizontal and vertical axes show the R.A. ($\Delta \alpha \rm \cos \delta$) and declination ($\Delta
\delta$) offsets in arcsec relative to 
$\alpha=06^{\rm h}$ 17$^{\rm m}$ 44.$^{\rm \!\! s}$2, 
$\delta=+22^{\rm o}$ 21\amin\ 49\asec\ (J2000).}  The white lines demark the edges of the regions mapped in each band.  The black circles centered at offset (--24$^{\prime\prime}$, --20$^{\prime\prime}$) indicate the regions for which the average spectra shown in Figure 1 were obtained.

\end{figure}
\clearpage

\begin{figure}
\includegraphics[scale=0.80,angle=0]{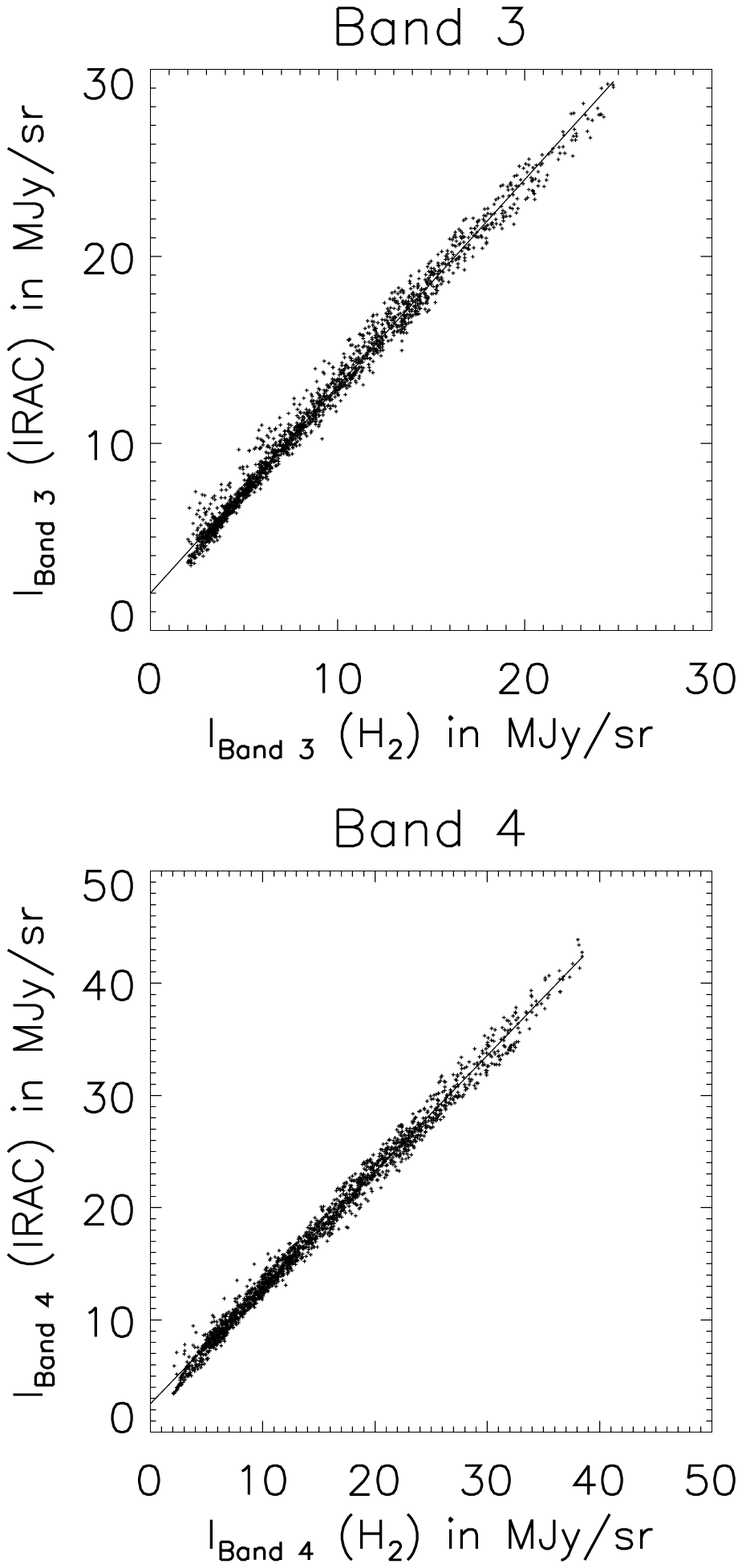}

\noindent{Fig.\ 4 -- Scatter plots comparing the IRAC-measured intensities , $I_{\rm Band \,3}({\rm IRAC})$ and $I_{\rm Band \,4}({\rm IRAC})$ on the vertical axis, with the intensities $I_{\rm Band \,3}({\rm H_2})$ and $I_{\rm Band \,4}({\rm H_2})$ expected given the IRS-measured spectral line maps for H$_2$ S(4) -- S(7).  Each point corresponds to a single $1.2^{\prime\prime} \times 1.2^{\prime\prime}$ pixel.   Solid lines show the result of a linear regression.  The slopes on the best-fit linear relationships are 1.10 and 1.03 respectively for Bands 3 and 4, and the y-intercepts are 2.0 and 2.5 MJy~sr$^{-1}$.}
\end{figure}

\begin{figure}
\includegraphics[scale=0.80,angle=0]{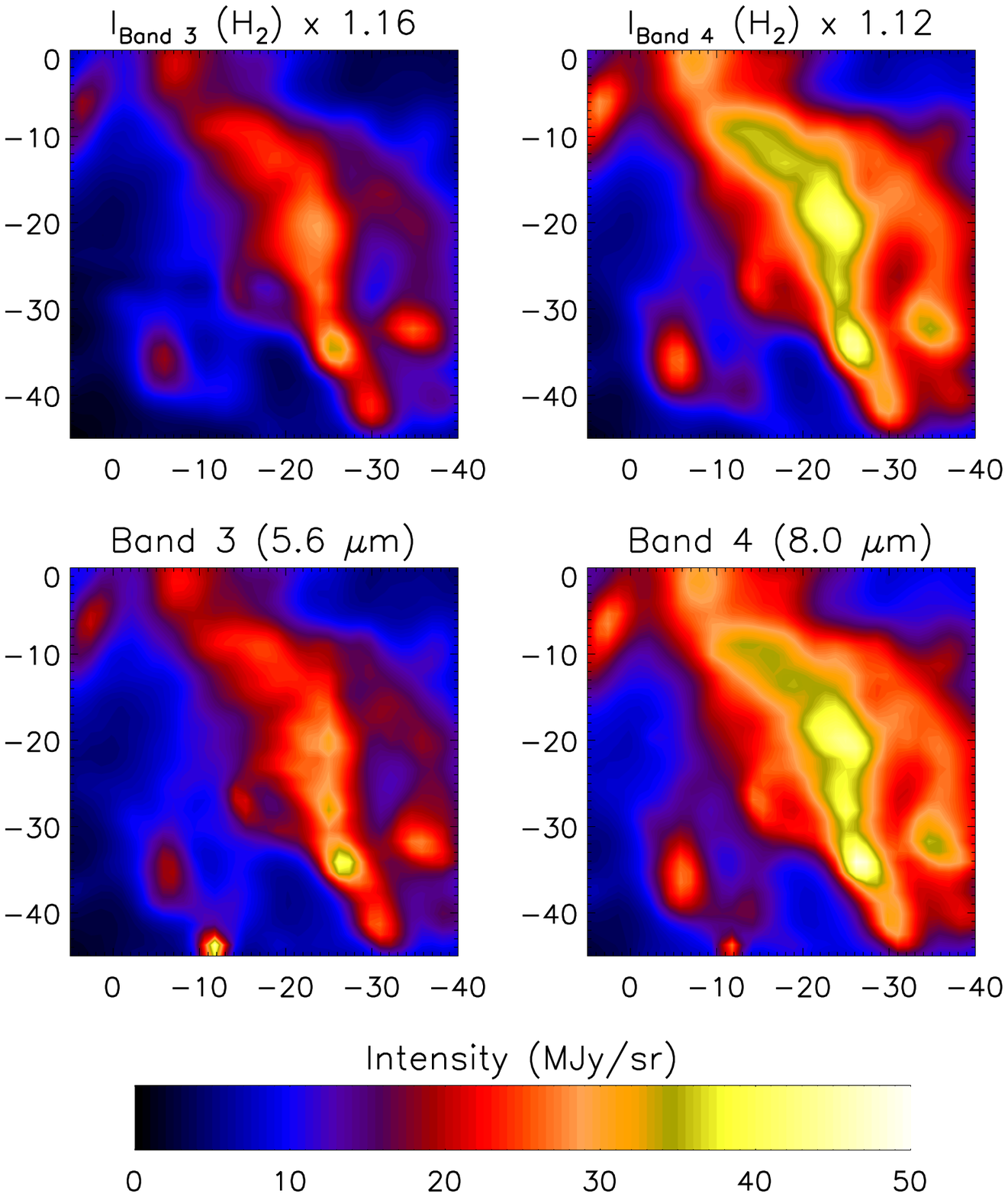}

\noindent{Fig.\ 5 -- Maps comparing the IRAC-measured intensities , $I_{\rm Band \,3}({\rm IRAC})$ and $I_{\rm Band \,4}({\rm IRAC})$, with the intensities $I_{\rm Band \,3}({\rm H_2})$ and $I_{\rm Band \,4}({\rm H_2})$ expected given the IRS-measured spectral lines maps for H$_2$ S(4) -- S(7).  \re{The horizontal and vertical axes show the R.A. ($\Delta \alpha \rm \cos \delta$) and declination ($\Delta
\delta$) offsets in arcsec relative to 
$\alpha=06^{\rm h}$ 17$^{\rm m}$ 44.$^{\rm \!\! s}$2, 
$\delta=+22^{\rm o}$ 21\amin\ 49\asec\ (J2000).}}
\end{figure}

\begin{figure}
\includegraphics[scale=0.60,angle=0]{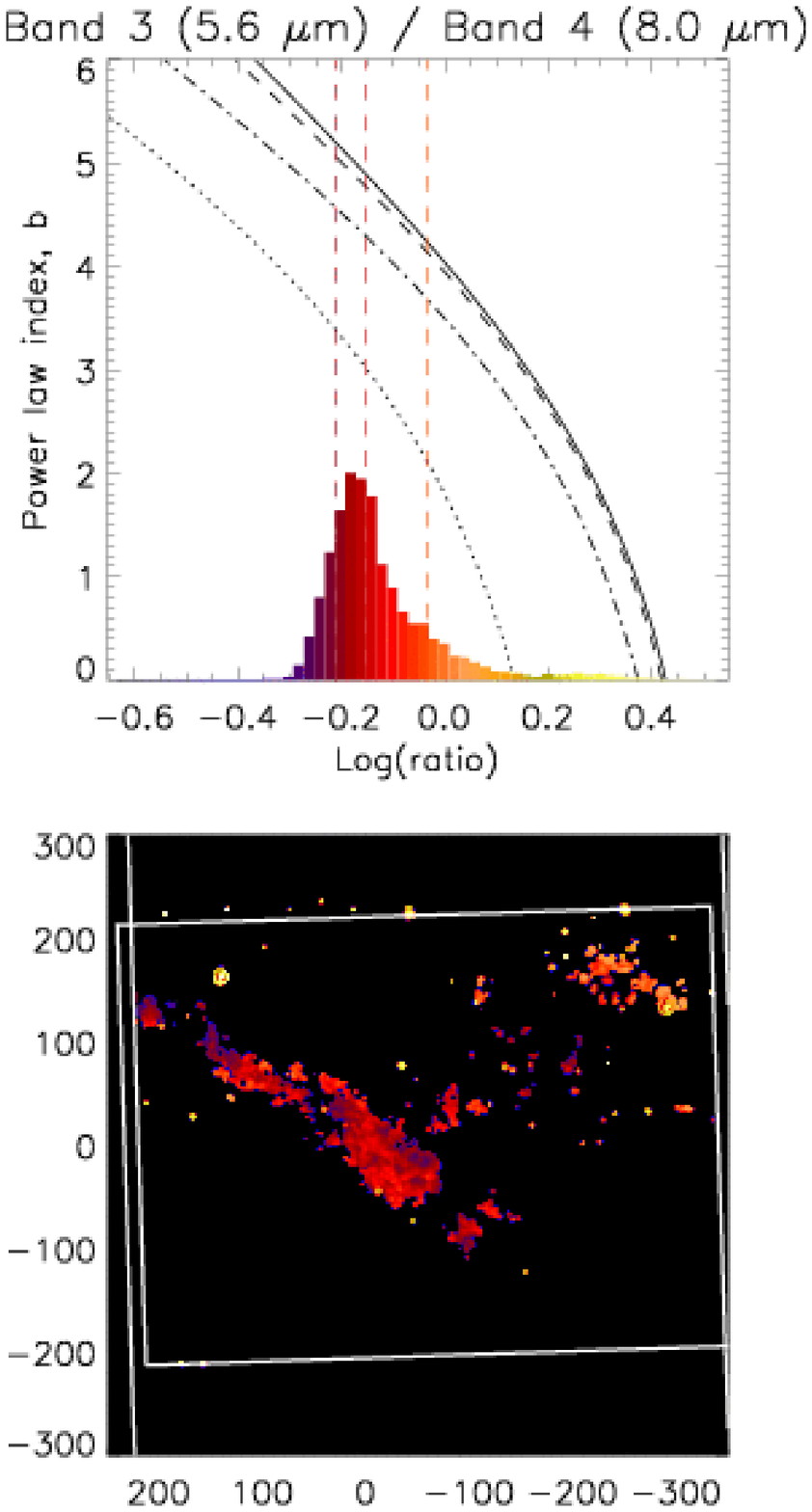}

\noindent{Fig.\ 6 -- Lower panel: map of the median-subtracted IRAC Band 3 / Band 4 ratio in IC443. The
horizontal and vertical axes show the R.A. ($\Delta \alpha \rm \cos \delta$) and declination ($\Delta
\delta$) offsets in arcsec relative to 
$\alpha=06^{\rm h}$ 17$^{\rm m}$ 44.$^{\rm \!\! s}$2, 
$\delta=+22^{\rm o}$ 21\amin\ 49\asec\ (J2000).
Regions in which the IRAC intensities are small (with a median-subtracted Band 3 intensity less than 3~MJy~sr$^{-1}$) are suppressed and appear in black. The white lines demark the edges of the regions mapped in each band.  Upper panel: a histogram indicates the relative number of pixels for which the logarithm of the band ratio lies in a given range (arbitrary scaling).  The colors appearing underneath the histogram curve are matched to those appearing in the lower panel, so the histogram serves as a color bar for the band ratio maps.  Vertical dashed lines indicate the values of log(band ratio) corresponding to the lower sextile, median, and upper sextile of the distribution.  Black curves indicate how the expected band ratio (plotted logarithmically on the {\it horizontal axis}) depends upon the power-law index, $b$ (appearing on the vertical axis and defined in \S4 below); results are given for H$_2$ densities of 10$^4$ (dotted curve), 10$^5$ (dot-dashed curve), 10$^6$ (dashed curve) and $10^7 \rm \, cm^{-3}$ (solid curve).}

\end{figure}

\begin{figure}
\includegraphics[scale=0.80,angle=0]{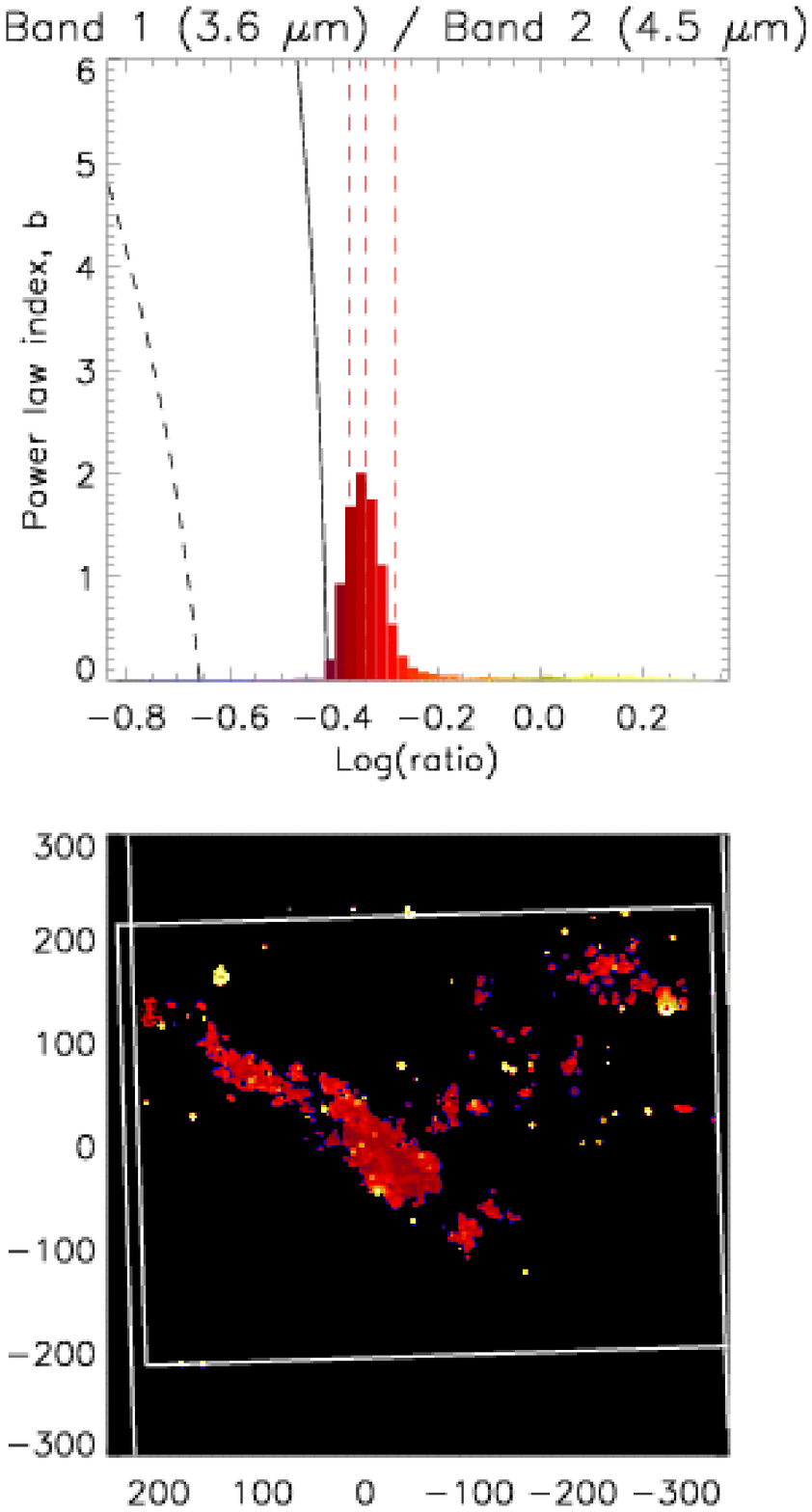}

\noindent{Fig.\ 7 -- Same as Fig.\ 6, except for the Band 1 / Band 2 ratio} 
\end{figure}

\begin{figure}
\includegraphics[scale=0.80,angle=0]{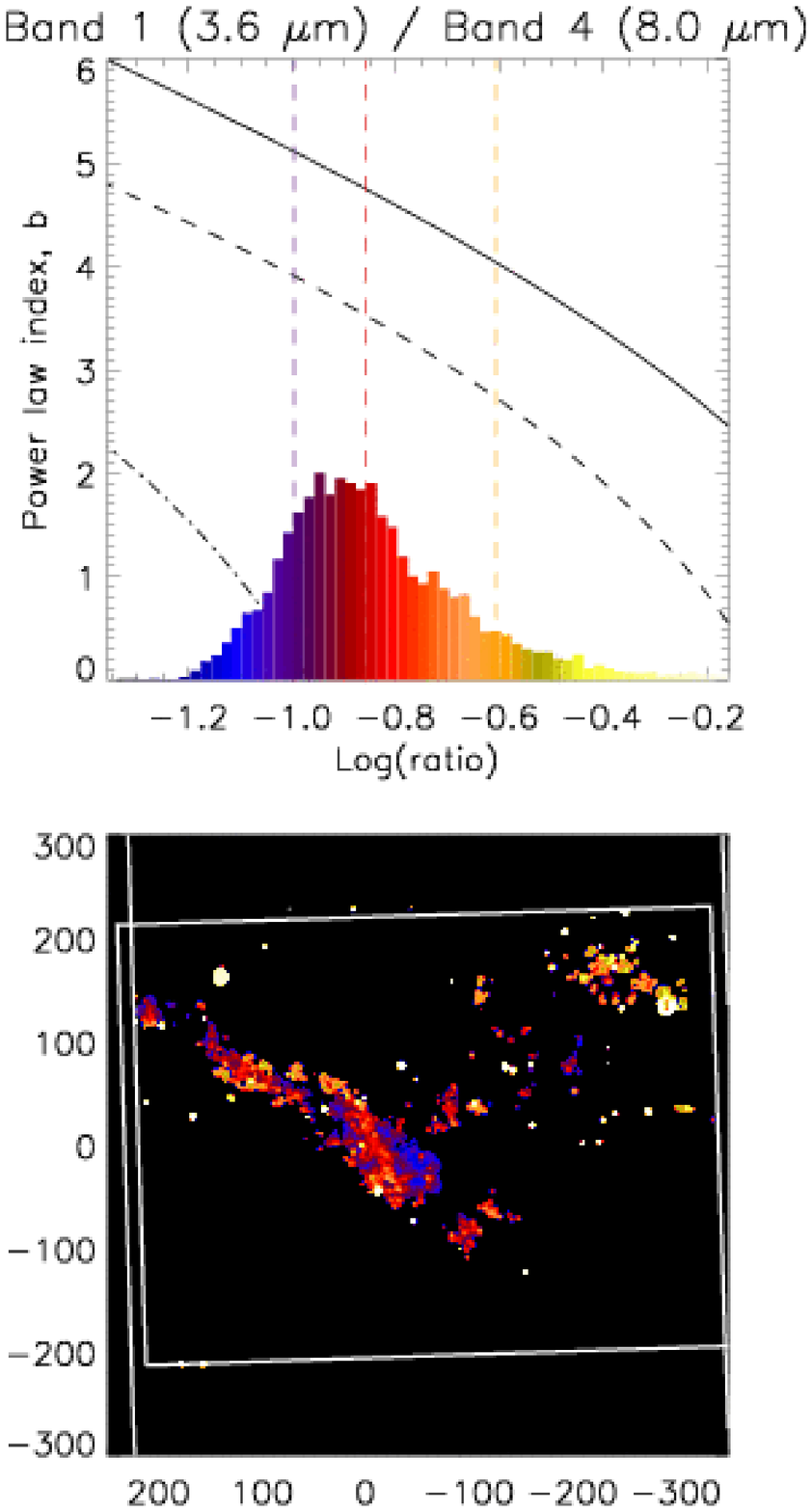}

\noindent{Fig.\ 8 -- Same as Fig.\ 6, except for the Band 1 / Band 4 ratio} 
\end{figure}

\begin{figure}
\includegraphics[scale=0.60,angle=0]{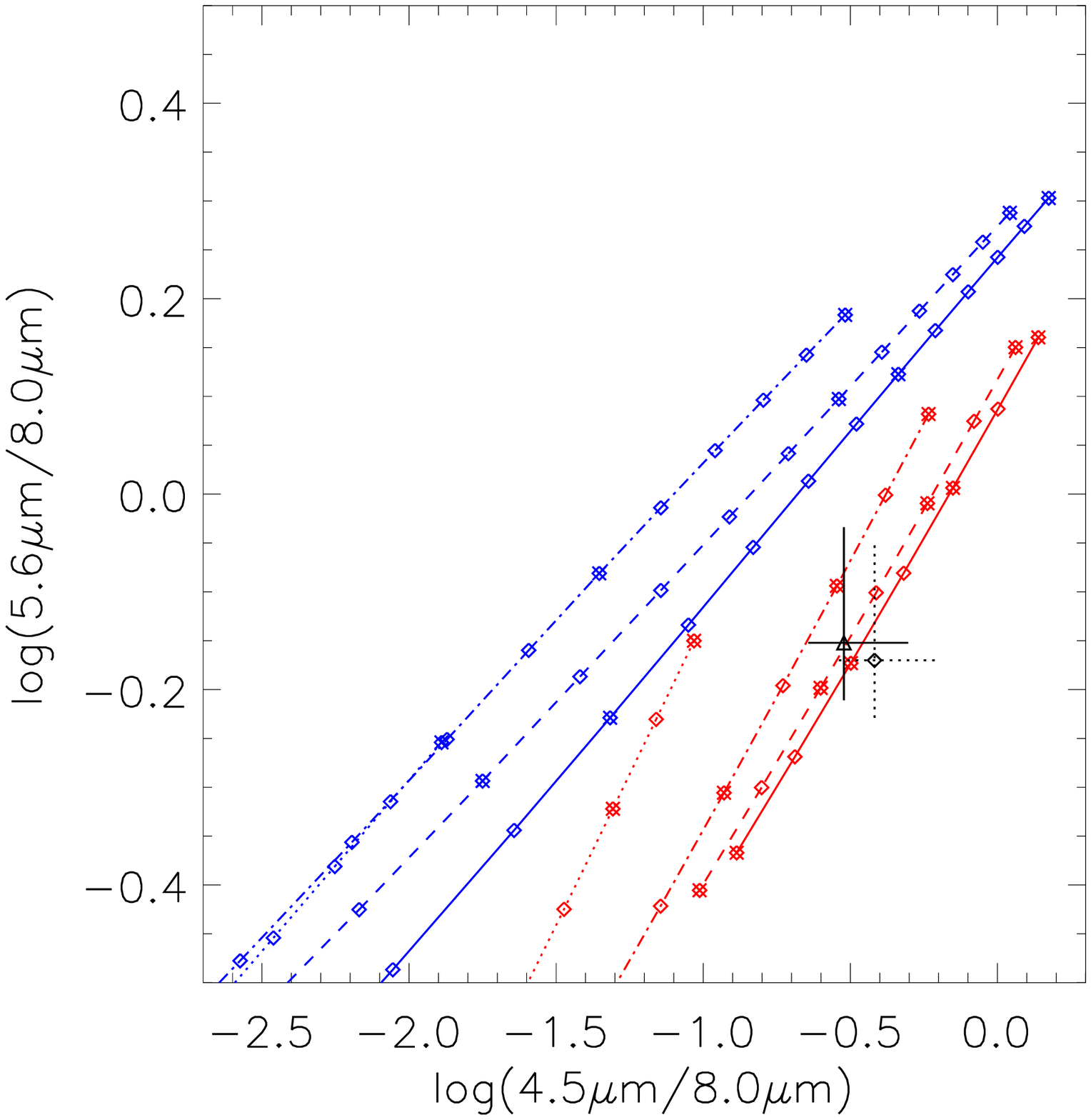}

\noindent{Fig.\ 9 -- Two-color diagram showing the Band 2 (4.5$\mu$m) / Band 4 (8.0 $\mu$m) and Band 3 (5.6$\mu$m) / Band 4 (8.0 $\mu$m) ratios expected for pure H$_2$ emission with an assumed visual extinction $A_V = 13.5$~mag.  Blue curves: single component models for H$_2$ densities of 10$^4$ (dotted curve), 10$^5$ (dot-dashed curve), 10$^6$ (dashed curve) and $10^7 \rm \, cm^{-3}$ (solid curve).  Results are shown for a range of temperatures up to 2000~K (top right of each curve), with \re{squares} every 100~K 
and \re{crosses} every 500~K.
The open triangle represents the median band ratios observed in the IRAC maps, with the error bars designating the upper and lower sextiles, all obtained with the standard point source flux calibration that we consider appropriate for this application (see discussion in \S 2 above). For comparison, the open diamond with the dotted error bars is obtained for flux calibration appropriate to a highly extended source. 
Red curves: models in which a power-law temperature distribution is assumed, with $dN = aT^{-b} dT$ (see eqn.\ 3) and $b$ in the range 3 (top right of each curve) to 6.  Results are shown for H$_2$ densities of 10$^4$ (dotted curve), 10$^5$ (dot-dashed curve), 10$^6$ (dashed curve) and $10^7 \rm \, cm^{-3}$ (solid curve). \re{Squares} appear when $b$ is an integral multiple of one-half, and \re{crosses} where $b$ is an integer.} 

\end{figure}

\begin{figure}
\includegraphics[scale=0.70,angle=0]{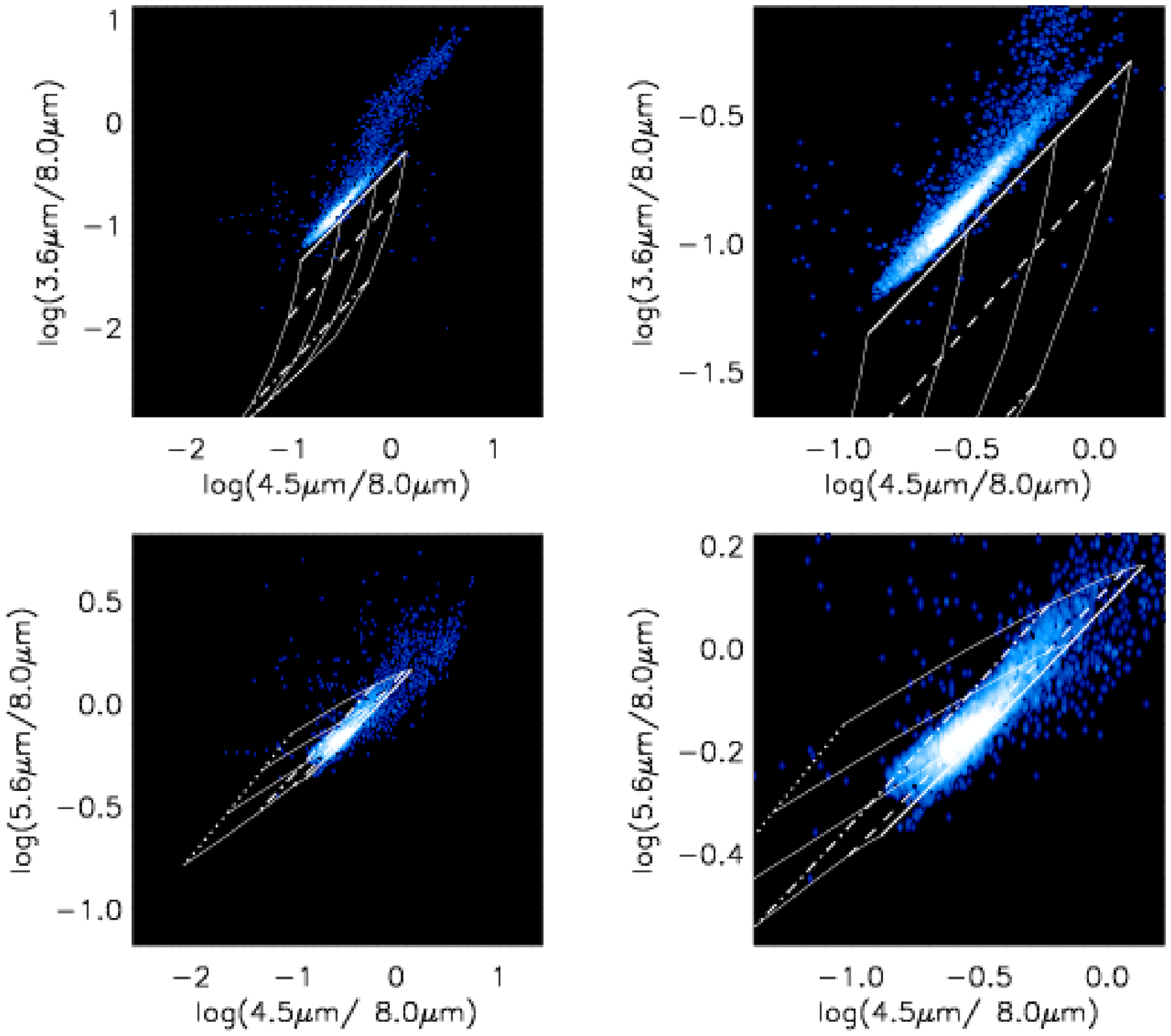}

\noindent{Fig.\ 10 -- Color-color diagrams for a selection of two pairs of band ratios, showing the two-dimensional distribution function for the logarithm of the observed band ratios.  The false color image (coded from black [smallest value] to blue to white [largest value]) represents the number of pixels for which the band ratios show a given pair of values.  Heavy white curves show the ratios expected for H$_2$ emission from gas with a power-law distribution of temperatures.  Results are shown for H$_2$ densities of 10$^4$ (dotted curve), 10$^5$ (dot-dashed curve), 10$^6$ (dashed curve) and $10^7 \rm \, cm^{-3}$ (solid curve).  Light white curves show the loci for power-law indices, $b$, of 3 (rightmost curve in top panel; uppermost curve in bottom panel), 4, 5, and 6. The right panels are zoomed versions of the left panels.}

\end{figure}

\begin{figure}
\includegraphics[scale=0.70,angle=0]{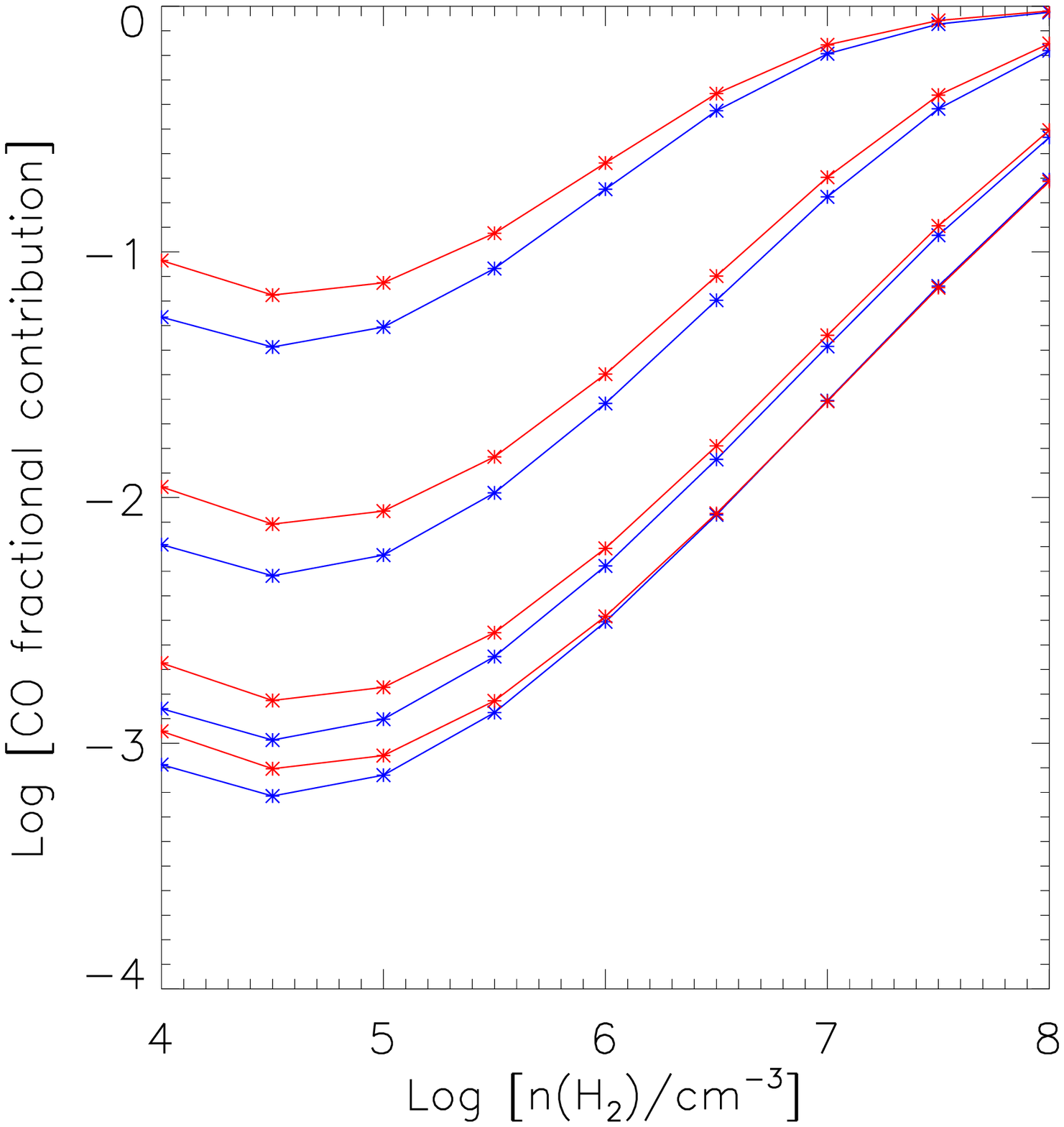}

\noindent\re{Fig.\ 11 -- Fractional contribution of CO emissions to IRAC Band 2, as a function of H$_2$ density and for an assumed CO/H$_2$ abundance ratio of $10^{-4}$.  Results are given for H/H$_2$ abundance ratios of 0, 0.01, 0.1, and 1 (from bottom to top), and for temperature power-law indices, $b$, of 4 (blue curves) and 5 (red curves).  }

\end{figure}

\end{document}